\documentclass[final,times,5p,twocolumn]{elsarticle}

\pdfpagewidth=\paperwidth
\pdfpageheight=\paperheight

\usepackage{natbib}
\usepackage{amsmath,graphicx}
\usepackage{color}
\usepackage{amsfonts,amsbsy,amssymb}
\usepackage[mathlines]{lineno}

\def\H{\mathcal H}
\def\e{{\rm e}}

\def\one{\mathbb I}
\def\z{\hat z}
\def\y{\hat y}
\def\x{\hat x}
\def\bq{\boldsymbol {\rm q}}

\begin{document}

\begin{frontmatter}

\title{Estimating yield-strain via deformation-recovery simulations}
\tnotetext[mytitlenote]{This work is a contribution of the National Institute of Standards and Technology and is not subject to copyright in the United States.}

\author[nist]{Paul N. Patrone \corref{corresponding}}
\ead{paul.patrone@nist.gov}
\cortext[corresponding]{Corresponding author}
\address[nist]{National Institute of Standards and Technology}
\author[boeing]{Samuel Tucker}
\address[boeing]{The Boeing Company}
\author[nist]{Andrew Dienstfrey}
\begin{abstract}
In computational materials science, predicting the yield strain of crosslinked polymers remains a challenging task.  A common approach is to identify yield as the first critical point of stress-strain curves simulated by molecular dynamics (MD).  However, in such cases the underlying data can be excessively noisy, making it difficult to extract meaningful results.  In this work, we propose an alternate method for identifying yield on the basis of deformation-recovery simulations.  Notably, the corresponding raw data (i.e.\ residual strains) produce a sharper signal for yield via a transition in their global behavior. We analyze this transition by non-linear regression of computational data to a hyperbolic model.  As part of this analysis, we also propose uncertainty quantification techniques for assessing when and to what extent the simulated data is informative of yield.  Moreover, we show how the method directly tests for yield via the onset of permanent deformation and discuss recent experimental results, which compare favorably with our predictions.
\end{abstract}

\begin{keyword}
 Yield strain \sep Molecular Dynamics \sep Crosslinked polymers \sep Uncertainty Quantification
\end{keyword}

\end{frontmatter}

\section{Introduction}
\label{sec:intro}

In computational materials science, estimating the ultimate mechanical properties of crosslinked polymers remains a challenging task \cite{Strachan15,economics}.  From an industrial, materials-design perspective, one of the key problems amounts to an inherent competition between throughput and model fidelity.  That is, on the one hand, the microscopic events that determine bulk properties involve many-body interactions over long distances; thus large, fully atomistic (or even quantum) simulations are needed to capture the relevant physics \cite{Binder02,berne98,deu99}.  On the other hand, such computations become prohibitively expensive long before the system size approaches the thermodynamic limit.  As a compromise, it is therefore becoming common for modelers to use atomistic molecular dynamics (MD) simulations of modest-sized systems\footnote{i.e.\ systems that model $\mathcal O(10^3)$ to $\mathcal O(10^4)$ atoms over tens of nanoseconds.} in the hopes that the corresponding predictions will {nevertheless} approximate bulk properties \cite{Strachan15,Binder02,Li11,Patrone16,Li15}.

\begin{figure}
\includegraphics[width=8cm]{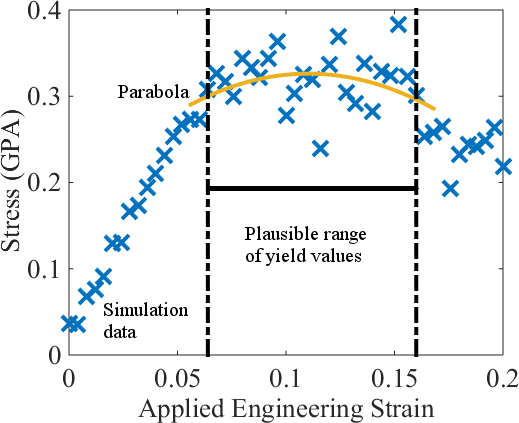}
\caption{An illustration of possible problems when extracting the yield strain $\epsilon_y$ from a stress-strain curve generated by MD.  The underlying system is a roughly 5000 atom unit cell composed of 33DDS and MY720 in a 1-to-1 ratio; see Sec.~\ref{subsec:systems} for description of this chemistry.  The polymers are crosslinked to roughly 90\%.  The noise is so large that a plausible window for yield ranges between $\epsilon = 0.06$ and $\epsilon = 0.16$, which may be useless for predictive purposes. To generate this estimate, we (i) picked a collection of 5 adjacent datapoints near the middle of the flat region, corresponding to $\epsilon \approx 0.11$; (ii) fit a constant to the data; and (iii) iteratively added neighboring points to this collection and recomputed the constant fit until the mean of the squared residuals began to significantly increase.  Thus, we interpret the plausible range of yield values as the window beyond which the data is not statistically constant.  To check the consistency of this estimate, we also fit a parabola to this domain extended by four additional datapoints, two to the left and two to the right.  According to this fit, $\sigma''(\epsilon_y) \approx -12$ GPA.  {Informally estimating the noise to be} $\varsigma \approx 0.05$ GPA yields an uncertainty estimate $\delta \approx \sqrt{-\varsigma / \sigma''(\epsilon_y)} \approx \pm 0.065$, which is visually consistent with the data and the plausible range of yields.  See the main text for a justification of this latter estimate.  }\label{fig:stressstrain}
\vspace{3mm}
\includegraphics[width=8.5cm]{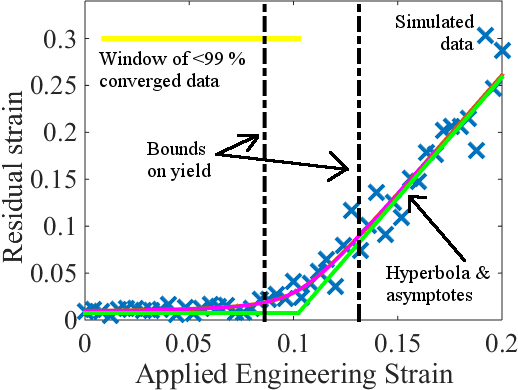}\caption{Residual strain as a function of applied engineering strain.  The underlying system is the same as in Fig.~\ref{fig:stressstrain}.  Note that in contrast with the former, there is a relatively sharp transition around $\epsilon \approx 0.1$ that is indicative of the material having yielded.  The analysis underlying the hyperbola fit (pink) and bounds on yield is described in Sec.~\ref{sec:analysis}.  The horizontal yellow line indicates the domain to the left of which all of the data is 99 \% or more converged to the lower asymptote (green).}\label{fig:residdata}
\end{figure}

While this trend is furthering the development of simulations as an industrial engineering tool \cite{economics}, it has also increasingly brought scientists into contact with the limitations of MD.  In the case of yield-strain $\epsilon_y$, such limitations typically manifest in the form of noise that complicates data analysis \cite{Strachan15,Li11,Li15,Hossain10,Li2014}.  In more detail, modelers often identify yield as the first local maximum of the von-Mises stress-strain curve $\sigma(\epsilon)$ \cite{Strachan2012,ref1,Hart10,Gosse01}.  However, finite-size simulations and limited time averaging allow for the motion of individual monomers and torsions to manifest as kinks in the resulting datasets, introducing artificial extrema (cf. Fig.~\ref{fig:stressstrain} and Refs.~\cite{Li11,Li2014,Li15,Veera13}).  Moreover, the spectrum of relaxation times for crosslinked polymers is poorly sampled by MD,  and hence it is not even obvious that simulated stress-strain curves necessarily capture yield processes.  As a result, modelers require improved simulation techniques and uncertainty quantification (UQ) analyses to better assess when and to what extent their predictions are informative of yield \cite{economics,Patrone16,Dienstfrey14}.

In this work, our goal is to address these problems by proposing a modified procedure for estimating $\epsilon_y$ on the basis of deformation-recovery simulations. {In place of stress-strain data, we analyze simulated residual strains $\epsilon_r$ as a function of the applied engineering strain $\epsilon$.  With respect to such data, yield is identified as the onset of permanent deformation, i.e.\ non-zero values of $\epsilon_r$.  As we show, this approach provides a more precise estimation of $\epsilon_y$ via a sharp transition in the behavior of the underlying raw data.  Importantly, this transition can be analyzed in terms of known techniques based on hyperbola fits, which are straightforward to implement and admit simple UQ techniques.  \textcolor{black}{Moreover, corresponding experimental results also indicate that the value of yield extracted from such a procedure is independent of the time over which the system is allowed to relax, suggesting that our approach is, to a certain extent, unaffected by the timescale limitations of MD.}

A key motivation for our procedure arises from the observation that analyses based on $\sigma(\epsilon)$ may lead to unfounded estimates of $\epsilon_y$ and/or unacceptable levels of uncertainty in its value.  {A major contribution to this uncertainty arises from the fact that the traditional, critical point estimates of $\epsilon_y$ depend on} a {\it local} property of {the stress-strain curve},\footnote{That is, a (non-global) critical point $\hat x$ is only associated with the behavior (i.e.\ zero slope) of a function $f(x)$ at a single point, namely $\hat x$.}  so that without additional outside information, most of the simulated data does not inform the value of yield.  
In such circumstances, the uncertainty $\delta$ in $\epsilon_y$ is roughly related to the noise $\varsigma$ via the approximation $\delta = \mathcal O\left(\sqrt{-\varsigma/\sigma''(\epsilon_y)}\right)$, which can be quite large for typical simulations (cf.\ Fig.~\ref{fig:stressstrain}).\footnote{This approximation is derived as follows.  First, let $\sigma'$ and $\sigma''$ denote the first and second derivatives of stress.  Next note that by definition, $\sigma'(\epsilon_y)=0$ and $\sigma''(\epsilon_y)<0$.  Therefore, the approximate value of $\Delta \epsilon = \delta$ required to see a change $-\varsigma$ in the stress is given by inverting $-\varsigma = (1/2) \delta^2 \sigma''(\epsilon_y)$, i.e.\ $\delta = \sqrt{-2\varsigma/\sigma''(\epsilon_y)}$. }  In order to overcome this issue, our analysis extracts $\epsilon_y$ from the {\it global behavior} of $\epsilon_r(\epsilon)$, which is characterized in terms of hyperbola asymptotes.  Importantly, this approach sharpens estimates of $\epsilon_y$ by using {\it all} of the available datapoints to stabilize the fitting process.  Moreover, the fit can itself be analytically interrogated to assess when a given dataset is too noisy to determine $\epsilon_y$.

A second motivation for our procedure stems from the desire to formulate an objective, reproducible, and automated workflow for estimating yield.  In the context of analyses based on stress-strain relations, this task remains challenging, given that there is no known, global functional form for $\sigma(\epsilon)$ describing crosslinked polymers in the nonlinear regime \cite{Perez98,Stach97}.  As a result, the associated methods for estimating $\epsilon_y$ necessarily vary among modelers and introduce uncertainty via subjective interpretations of the data.  In contrast, our method relies on the generic observation that $\epsilon_r=0$ and $\epsilon_r \propto \epsilon-\epsilon_y$ when the applied deformations are below and above yield, respectively.  As this behavior is universal across a  range of material systems, our corresponding data analysis can be applied in a uniform and reproducible manner with little-to-no input from the modeler.  

Despite these stated aims, however, we emphasize that the scope of our work is limited to an analysis of yield (and its uncertainty) {\it within} the context of a single simulation.  This has important implications insofar as a single, small simulation may not capture all sources of uncertainty \cite{Patrone16}.  In the case of the glass transition temperature $T_g$, Ref.~\cite{Patrone16}  addressed this problem via a weighted-mean average designed to account for the uncertainty between multiple simulated datasets.  However, this estimate relied on a prerequisite convergence analysis to help verify that it was indeed appropriate to average datasets in the first place.  Given that it is not clear how to extend this convergence test to the case of yield, we refrain from invoking a similar between-simulation uncertainty analysis.  Here our goal is simply to propose and justify a modified simulation procedure for estimating $\epsilon_y$, and we leave a final assessment of the total simulated uncertainties for future work.  

Along related lines, we do not exhaustively {\it validate} our method against experimental results.  Recent work has demonstrated that strain-recovery can be used to estimate yield of glassy polymers in laboratory settings \cite{Jackson16}, and we show how our method makes qualitatively similar predictions.  However, a more quantitative comparison requires extensive estimation of model-form uncertainties.  Given that {\it verification} (or estimation of uncertainties within the computational realm) remains an open problem, we postpone rigorous validation until the former task is more fully addressed.  Such tasks comprise ongoing studies.

The rest of this paper is organized as follows.  In Sec.~\ref{sec:simproc}, we discuss the simulation procedure needed to generate residual-strain data.  In Sec.~\ref{sec:analysis}, we discuss our method for extracting $\epsilon_y$ and the statistical analyses underlying the associated UQ.  Section \ref{sec:discussion} further discusses our method in the context of stress-strain curves and experimental data, and appendices provide key mathematical ideas underlying the simulation protocol.

\section{Simulation procedure}
\label{sec:simproc}

\subsection{Procedure Overview}

Our simulation protocol is implemented via custom Materials Studio\textsuperscript{\texttrademark} scripts and consists of three main steps: (i) crosslinking and annealing; (ii) iterative straining; and (iii) strain relaxation.\footnote{Certain commercial equipment, instruments, or materials are identified in this paper in order to specify the experimental procedure adequately. Such identification is not intended to imply recommendation or endorsement by the National Institute of Standards and Technology, nor is it intended to imply that the materials or equipment identified are necessarily the best available for the purpose.}  As the first of these steps has been documented elsewhere \cite{Patrone16}, we only review key points.  The second and third steps are discussed in more detail below.

Starting with an unreacted unit cell of monomers, we crosslink the system by iterating between nearest-neighbor bonding steps and short relaxation simulations that eliminate high-energy configurations.  After the monomers are sufficiently crosslinked (between 60\% and 95\%), the system is heated to 800 K and then annealed to 300 K in 10 K increments using the Parrinello barostat \cite{Parrinello81} and Andersen thermostat \cite{Andersen80}. A convergence criterion on the density adaptively determines the number of timesteps at each temperature.  The criterion first allows the density to equilibrate to its new temperature until fluctuations in its running average fall below some threshold.  Then, this running average is discarded, and a new running average density is computed until its fluctuations fall below a second, stricter threshold.  After this criterion is met, the temperature is decreased and the procedure repeated; cf. \ref{app:convergence}.  Once the system has reached 300 K, the running average cell dimensions of the last averaging step are taken to be the dimensions of the relaxed, unstrained system. 

Given an annealed structure, we next apply a small, volume conserving strain $\Delta \epsilon$ (i.e.\ Poisson's ratio of 1/2) to the system by compressing in the direction of the longest basis vector.\footnote{As an aside, we emphasize that this method does not assess a possible anisotropic response of the material to different loading conditions.  Here, our method is chosen solely to ensure that none of the system dimensions become smaller than the electrostatic interaction cutoff distance.}  We apply a tensile stress in two other orthogonal directions.  Because the annealed structures are generally skew, this involves both lengthening the basis vectors and changing their relative angles; see  \ref{app:non-orthogonal} for details of the calculations.  Next, we equilibrate the system via an NVT simulation that adaptively chooses the number of timesteps according to a convergence criterion analogous to the one described previously, but {in this case monitoring the stress}.  As above, each iteration uses a second, more stringent convergence criterion to compute a final, average stress for use in subsequent analysis.  In addition, {the final structure obtained here is saved for later use to evaluate the residual strain as discussed below}.  In order to generate a full stress-strain curve, we increment the applied strain and iterate the above steps until enough datapoints $(\epsilon_i,\sigma_i)$ have been collected, typically 51 in total (i.e. corresponding to one zero-strain structure and $50$ strain increments).

After this process has completed, we allow each of the $N$ saved structures to relax using an NPT simulation.  As above, the number of timesteps is adaptively chosen until fluctuations in the running average system dimensions fall below a threshold, and the final average dimensions are computed using a second, more stringent criterion.  Denoting the length of the sides of the relaxed and original unit cells as $\ell_i$ and $L_i$, we define the residual strain as
\begin{align}
\epsilon_r := \sum_{i=1}^3 \left| \frac{L_i - \ell_i}{L_i} \right|.  \label{eq:resid}
\end{align}  
The absolute values in Eq.~\eqref{eq:resid} ensure that $\epsilon_r$ is largest when all three $\ell_i$ differ from the original lengths.  Our definition is consistent with those used in Refs. \cite{Jackson16,Pegoretti06,PEN97}, with the exception that we consider strain in all three directions as opposed to just the compression axis.

\subsection{Systems considered in this work}
\label{subsec:systems}
In this work, we consider two chemistries representative of materials found in aerospace applications.  The first of these, which we denote 33MY, is a one-to-one mixture of: 3,3-diaminodiphenyl sulfone (33DDS), a four-functional amine; and tetraglycidyl methylene dianiline (MY720), a four-functional epoxy.  The second system, which we denote 44BA, is a one-to-two mixture of: 4,4-diaminodiphenyl sulfone (44DDS); and digycidyl ether of Bisphenyl A (BisA), a two-functional epoxy.  The second chemistry was chosen because Ref.~\cite{Jackson16} recently performed experiments on this systems using techniques analogous to our strain-recovery simulations.  

All of our simulated results were generated from unit cells with roughly 4000 to 5000 atoms.  While this is small relative to state-of-the-art calculations \cite{Strachan15,Li11,Li15,Strachan2012}, our emphasis is on generating data that is representative of what might be encountered in a high-throughput environment.  Sizes above roughly 10,000 atoms can take weeks to months to simulate on high-end desktop machines, which may be prohibitive from a cost perspective (cf.\ also Ref.~\cite{Patrone16}).  Moreover, our goal is not to generate the most well-behaved data, but rather to show that reasonable data requires thoughtful analysis.  To that end, we chose chemistries and system sizes to demonstrate a range of possible behaviors encountered in practice.  

\section{Data analysis}
\label{sec:analysis}

\subsection{Observations on residual strain data}
Figure \ref{fig:residdata}, shows a representative example of a residual strains computed from the system that  generated Fig.~\ref{fig:stressstrain}.  Before proposing our data analysis routine, several comments are in order.

First, $\epsilon_r(\epsilon)$ has bilinear behavior that admits a simple physical interpretation.  When the applied strain is below yield, the system is able to fully recover to its original dimensions; thus the corresponding $\epsilon_r$ are close to zero and constant.  When the applied strain is greater than yield, only the pre-yield deformation is recoverable, leading to an approximately linear growth of $\epsilon_r$ with $\epsilon$.  Thus, in principle, $\epsilon_y$ can be identified as the value of applied strain at which the linear growth first appears.  

In practical simulations, such transitions are not sharp; we attribute this to several effects.  For one, strain-relaxation in polymers occurs via two distinct mechanism, namely elastic and viscoelastic recovery \cite{ward2004}.  The former occurs instantaneously and is associated with deformation modes that are confined to local energy minima, such as small changes in bond lengths.  Viscoelastic effects, on the other hand, are typically associated with transitions between energy minima, such as the torsional rearrangement between discrete states, collective sliding of chains, and so forth.  As the barriers between such states may be significant, the associated relaxation modes occur over physical timescales that may be unaccessible to MD.  Therefore, it is reasonable to expect that simulated $\epsilon_r(\epsilon)$ will smoothly transition from 0 to a post-yield, linear behavior over some finite interval of applied strains.\footnote{The convergence criterion in \ref{app:convergence} is designed to alleviate this problem.  However, the underlying calculation considers running averages computed on finite (and typically small) blocks of time, often on the order of 10 ps.  As a result, relaxation modes with significantly longer characteristic times are not adequately sampled by the simulations.  }

Thermal noise and finite-size effects also complicate our interpretation of the data.  In particular, at small applied strains one expects that $\epsilon_r = 0$.  However, fluctuating system dimensions conspire with the absolute values in Eq.~\eqref{eq:resid} to return small, non-zero $\epsilon_r$.  Beyond yield, the data in Fig.~\ref{fig:residdata} also shows a marked {\it increase} in fluctuations about linear behavior.  In the context of stress-strain curves, Sundararaghavan and Kumar \citep{Veera13} provided convincing evidence that such large fluctuations are due to a finite number of torsional rearrangements and related transitions that become activated post-yield.  By virtue of  the small system size, such rearrangements produced noticeable perturbations in their data.  Given the similarity to systems that we consider in this work, we anticipate that the increase in our post-yield noise arises for similar reasons.  Below we pay careful attention to ensure that our estimates of $\epsilon_y$ are not overly affected by such effects.

Along related lines, we also note that in several regards, Fig.~\ref{fig:residdata} is visually consistent with its underlying stress-strain curve in Fig.~\ref{fig:stressstrain}.  In particular, the transition region in the former corresponds roughly to a plausible location for the maximum of $\sigma(\epsilon)$.  Moreover, our general observations about the trends in noise apply equally to both.  By eye, however, the plausible region for $\epsilon_y$ is much smaller in Fig.~\ref{fig:residdata} than in Fig.~\ref{fig:stressstrain}.  In the following section, we propose analyses to make such observations more precise.  

\subsection{Hyperbola fits of residual strain}
\label{subsec:hyperbola}

As noted in the previous section, under ideal circumstances all of the applied strain up to yield is recoverable, while post-yield strain is not.  Mathematically this observation can be stated by letting $ \epsilon_r(\epsilon)$ be a function of the form
\begin{align}
\epsilon_r = a + b\Theta(\epsilon - \epsilon_y) [\epsilon - \epsilon_y], \label{eq:idealresid}
\end{align}
where $\Theta(x)$ is the Heaviside step function, and $a$ and $b$ are constants; cf.\ the green asymptotes in Fig.~\ref{fig:residdata}.  We emphasize that this function is at best an approximation for $\epsilon_r$.  \textcolor{black}{For one, there is (to the best of our knowledge) no general theory that predicts such a bilinear behavior for the residual strain, despite ample experimental evidence \cite{Jackson16,ref1,Perez98,Pegoretti06,PEN97}.  Moreover, we physically expect that $a=0$, but as has been noted practical simulations often return small, non-zero values of $\epsilon_r$ when $\epsilon \to 0$.    Nonetheless, the data shown in Fig.~\ref{fig:residdata} (along with subsequent figures) demonstrates that Eq.~\eqref{eq:idealresid} is a good  approximation to the behavior of $\epsilon_r(\epsilon)$.}

In order to account for the smoothing effects of finite-time and -size simulations, we fit $\epsilon_r$ to a hyperbola as follows.  Specifically, define
\begin{align}
\mathcal H =  a+ b (\epsilon - \epsilon_y)/2 + b \sqrt{(\epsilon - \epsilon_y)^2/4 + \e^{c}}, \label{eq:hyperbola}
\end{align}
where $a$, $b$, $c$, and $\epsilon_y$ are to be determined.  Note that when $c\to -\infty$, the hyperbola given by Eq.~\eqref{eq:hyperbola} {tends in the limit to the double line formula given by Eq.~\eqref{eq:idealresid}.  Finite $c$ allows for a smooth transition between the two linear asymptotes. }Equation~\eqref{eq:hyperbola} automatically enforces the requirement that below yield, the residual strain approaches a constant (ideally {$a\approx 0$}).   Given a set of parameters $\phi=(a,b,c,\epsilon_y)$ fit to a particular dataset, we equate the hyperbola center with yield $\epsilon_y$.

In order to actually determine $\phi$, we perform a weighted, non-linear least squares fit of the simulated data to Eq.~\eqref{eq:hyperbola} via
\begin{align}
\phi_{\rm LS} = {\rm argmin}_{\phi} \sum_{j=1}^D \left [\frac{\mathcal H(\phi,\epsilon_j) - \epsilon_{r,j}}{\varsigma_j} \right]^2 \label{eq:LS}
\end{align}
where $\epsilon_j$ and $\epsilon_{r,j}$ are the applied and residual strain values returned at the $D$ discrete simulation points, and $\varsigma_j$ is weighting factor associated with the amplitude of the noise in the simulations.  This latter quantity is particularly important in order to prevent the hyperbola from overfitting the increasingly noisy, high-strain data; see Fig.~\ref{fig:residdata}.  As we show below, failure to do so can lead to  unphysical values of yield.  

In order to estimate $\varsigma_j$, we use an iterative approach wherein we first assume $\varsigma_j=1$.  Solving the least-squares problem associated with Eq.~\eqref{eq:LS} provides an initial set of parameters $\hat \phi_0$ corresponding to a hyperbola $\H(\hat \phi_0,\epsilon)$.  Defining the squared residuals $r_j^2 = [\H(\hat \phi_0,\epsilon_j) - \epsilon_{r,j}]^2$, we fit these to a power law of the form
\begin{align}
R(\boldsymbol {\rm q},\epsilon) =q_1 + q_2 \epsilon^{q_3}. \label{eq:residmodel}
\end{align}
The {non-constant variance model \eqref{eq:residmodel}} is determined by maximizing the likelihood 
\begin{align}
\mathcal L = \prod_j \exp\left[\frac{-r_j^2}{q_1 + q_2 \epsilon_{j}^{q_3}} \right] \label{eq:likelihood}
\end{align}
 as a function of $\boldsymbol {\rm q}$.  Taking $\varsigma^2(\epsilon) = R(\boldsymbol {\rm q},\epsilon)$, we then compute a final estimate of the hyperbola parameters $\phi_{\rm LS}$ by solving Eq.~\eqref{eq:LS} with the updated weighting factor evaluated at the $\epsilon_j$.

Figure~\ref{fig:residdata} shows the results of this method for the residual strain experiments {corresponding to the computational stress-strain shown in} Fig.~\ref{fig:stressstrain}.  Figure~\ref{fig:residmodel} shows the {noise-model} fit according to Eq.~\eqref{eq:residmodel}. As a result of the iterative update for $\varsigma$, the hyperbola fits the small-strain data well while exhibiting some flexibility in its interpolation beyond $\epsilon_y$.  This is reasonable, given that the fluctuations in the former are small.  Figure~\ref{fig:scaledresids} also shows the residuals scaled by $R(\bq,\epsilon)$.  By eye, these appear {uniform in scale} and uncorrelated.  We take this as evidence that the {uncorrelated noise model with strain-dependent variance} \eqref{eq:residmodel} is sufficient to describe the associated fluctuations in $\epsilon_r$.

\begin{figure}
\includegraphics[width=8cm]{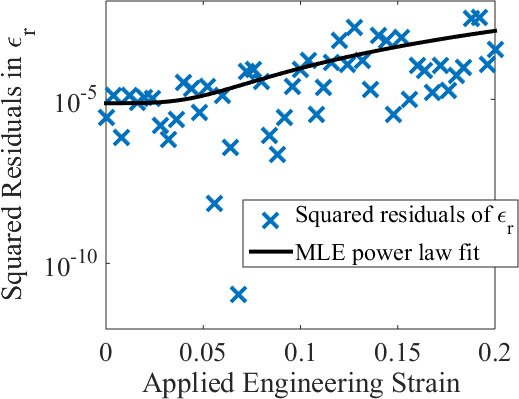}\caption{Power law fit of the squared residuals of $\epsilon_r$ according to the maximum-likelihood estimate (MLE) given by Eq.~\eqref{eq:residmodel}.}\label{fig:residmodel} 
\vspace{3mm}
\includegraphics[width=8cm]{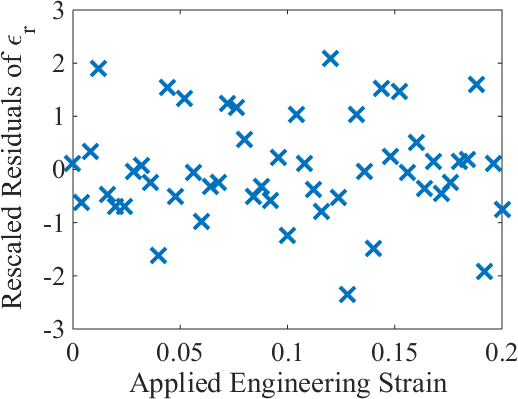}\caption{Residuals from the above figure rescaled according to the power law $R(\bq,\epsilon)$.  By eye, there is a lack of correlations between the data, suggesting that a white noise model sufficiently describes their randomness.}\label{fig:scaledresids}
\end{figure}

\subsection{Rejection criterion for datasets}

Considered as a tool for uncertainty quantification, Eq.~\eqref{eq:hyperbola} is useful for assessing when a given dataset is informative of yield.  In particular, the parameter $c$ can be used to quantify the extent to which the data approaches asymptotic regimes of the hyperbola.  Specifically, Ref.~\cite{Patrone16} showed that for a user-defined convergence threshold $P$ (where $0 < P < 1$), the formula
\begin{align}
\Delta \epsilon = \frac{e^{c/2}(2 P - 1)}{\sqrt{P(1-P)}} \label{eq:asymptotic}
\end{align}
defines an interval $[\epsilon_y-\Delta \epsilon,\epsilon_y + \Delta \epsilon]$ outside of which the $\epsilon_r$ are $100\times P$ \% converged to their nearest asymptote.  Thus, given a value of $P$, we can use Eq.~\eqref{eq:asymptotic} to identify those datasets for which the residual strain approaches a constant value when the applied strain is small.  Any simulation for which $\epsilon_y - \Delta \epsilon < 0$ can be held out for further inspection on the grounds that the data does not sample any deformation states that entirely recover.  

 In Fig.~\ref{fig:residdata}, we illustrate an analysis according to Eq.~\eqref{eq:asymptotic} applied to the data underlying Fig.~\ref{fig:stressstrain} with $P=0.99$.  The strictness of the convergence criterion is reflected in the fact that the residual strains essentially overlap the hyperbola asymptotes up to the region outlined in black, which was later identified as the confidence interval for $\epsilon_y$.

\subsection{Uncertainty quantification of $\epsilon_y$ through noise sampling}

As Fig.~\ref{fig:residdata} illustrates, noise in the simulated data leads to uncertainty with regards to the exact location of $\epsilon_y$.  By eye, it appears that $\epsilon \approx 0.08$ is a plausible lower bound, although such an assessment is largely subjective.  In this section, we propose a statistical method for estimating a confidence interval containing $\epsilon_y$ on the basis of repeated noise sampling.  Physically, our goal is to quantify the extent to which finite-size and -time fluctuations in the data inhibit our ability to compute a best-fit hyperbola.

Given $\hat \phi_0$ and $R(\bq,\epsilon)$ from Sec.~\ref{subsec:hyperbola}, we note that the latter amounts to a noise model for the residuals of $\epsilon_r$.  Letting $\mathcal N(0,x)$ denote a normal random variable with zero mean and variance $x$, we therefore construct synthetic datasets
\begin{align}
\hat \epsilon_{r,i} = \mathcal H(\hat \phi_0,\epsilon_{i}) + \mathcal N_i(0,R(\bq,\epsilon_i)) \label{eq:synthetic}
\end{align}
by using random number generators to realize the $\mathcal N_i$.  Fitting these synthetic datasets to hyperbolas via Eq.~\eqref{eq:LS} [with $\varsigma_i^2 = R(\bq,\epsilon_i)$] generates new estimates of $\epsilon_y$.  Given that this process is inexpensive, we realize $\mathcal O(10^5)$ or more such estimates and compute confidence intervals on the basis of the resulting distribution.  

Figures \ref{fig:yieldhist} and \ref{fig:newss} shows the results of this procedure applied to the simulation results that generated Figs.~\ref{fig:stressstrain} -- \ref{fig:residmodel}.  In Fig.~\ref{fig:newss}, we take the confidence intervals for yield (dash-dot black lines) to be the minimum and maximum values of $\epsilon_y$ returned by the noise sampling algorithm.  For the time being, we avoid reporting a mean and standard deviation since it is unclear whether the initial departure from $\epsilon_r \approx 0$ is due to the onset of yield or an inability to sample long-relaxation time viscoelastic modes.  Nonetheless, Figs.~\ref{fig:residdata} and \ref{fig:newss} show clearly that the residual strain method can lead to a marked decrease in the uncertainty in yield.  

\begin{figure}
\includegraphics[width=8cm]{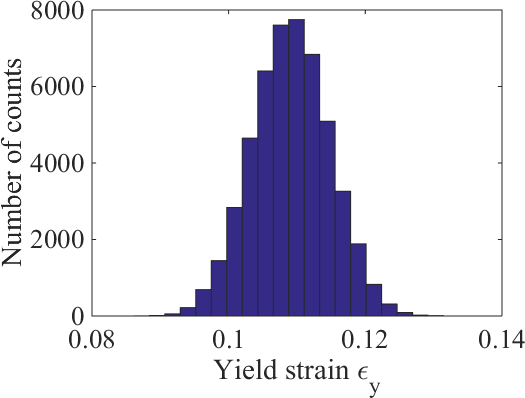}\caption{Histogram of plausible values of $\epsilon_y$ for the simulation considered in Figs.~\ref{fig:stressstrain}--\ref{fig:residmodel}.  Such estimates were obtained by generating 50,000 synthetic datasets according to Eq.~\eqref{eq:synthetic} and extracting yield from the corresponding hyperbola fits.}\label{fig:yieldhist}
\vspace{3mm}
\includegraphics[width=8cm]{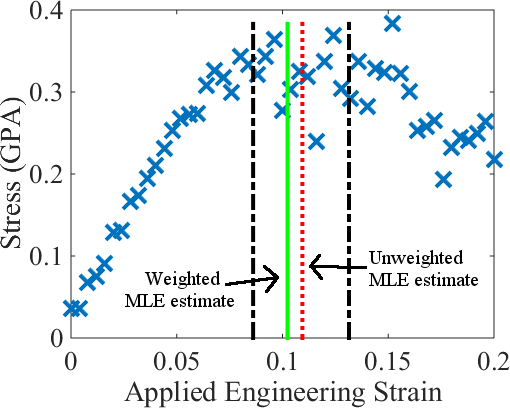}\caption{Stress-strain data in Fig.~\ref{fig:stressstrain} compared with a hyperbola analysis of the corresponding residual strains.  Note that the window of plausible strains (dash-dot black lines) is narrower than the region over which the data appears flat.  The vertical, dotted red line denotes the estimate of yield associated with taking $\varsigma=1$ in Eq.~\eqref{eq:LS}.  The vertical, solid green line is associated with the corresponding estimate when $\varsigma^2 = P(\boldsymbol {\rm q},\epsilon)$.   }\label{fig:newss}.
\end{figure}

\section{Discussion and conclusions}
\label{sec:discussion}

\subsection{Behaviors seen in data}
Amine-cured epoxies are a rich class of materials, and {\it in silico} they exhibit a variety of behaviors that affect the data analysis.  In this section, we discuss such considerations in more detail.  Relevant conclusions are summarized within each subsection.

\subsubsection{Overfitting and non-weighted least squares estimates}

In Sec.~\ref{subsec:hyperbola}, we suggested that a non-weighted least squares estimate of the parameters $\phi_0$ can lead to a hyperbola that overfits the residual strain data.  Figure~\ref{fig:faroff} shows an example of this situation.  The dashed, purple hyperbola corresponds to a non-weighted least squares fit with $\varsigma^2=1$.  At large applied strains, the fit follows the data significantly better than its weighted counterpart.  However, examination of the inset shows that the non-weighted least squares estimate of yield is unphysically high near a value of $\epsilon_y=0.19$.  We anticipate that this is due to overfitting the high-strain data, thus necessitating the iterative analysis by which we determine $R(\bq,\epsilon)$.  The weighted-least squares hyperbola (solid pink) allows for more flexibility in fitting the high-strain data, and leads to correspondingly more reasonable estimate of $\epsilon_y$.

\begin{figure}
\includegraphics[width=8cm]{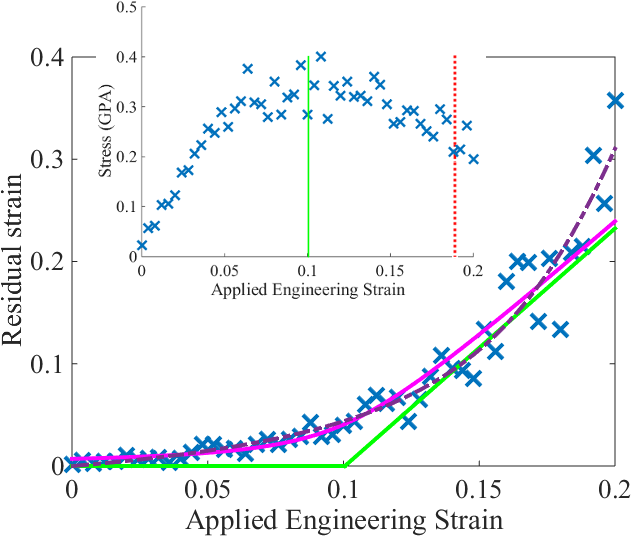}\caption{Comparison of hyperbolas computed via Eq.~\eqref{eq:LS} when $\varsigma^2 = 1$ (dashed purple) and $\varsigma^2=R(\bq,\epsilon)$ (pink with green asymptotes).  The inset shows the corresponding stress-strain data with the hyperbola-based yield estimates indicated by vertical lines; colors and linestyles have the same meaning as in Fig.~\ref{fig:newss}. This dataset comes from another 5000 atom 33MY simulation. }\label{fig:faroff}
\end{figure}

\subsubsection{Comparison to methods based on $\sigma(\epsilon)$}

Up to this point, we have suggested that the residual strain method provides estimates of $\epsilon_y$ that are both consistent with and sharper than their $\sigma(\epsilon)$ counterparts.  This is not always be true.  Figure \ref{fig:badpeak} illustrates a case for which the peak of the stress-strain data appears well resolved and localized around $\epsilon=0.1$, but outside the confidence interval determined from an analysis of 50,000 synthetic datasets.   Moreover, a plausible eyeball estimate of the uncertainty in $\epsilon_y$ extracted from $\sigma(\epsilon)$ is $\delta = \pm 0.01$, so that there is little overlap with the residual strain estimate.  Such a discrepancy could arise from a random fluctuation in $\sigma(\epsilon)$ that creates an artificial maximum.  However, this observation does not necessarily imply that our residual strain method provides a more accurate estimate of $\epsilon_y$ in this case.  Notably, the right-most bound from the latter method falls far into the post-yield region relative to the $\sigma(\epsilon)$ data.  Thus, this example highlights the potential need for estimates that combine information from multiple sources in order to overcome their individual limitations.  

\begin{figure}
\includegraphics[width=8cm]{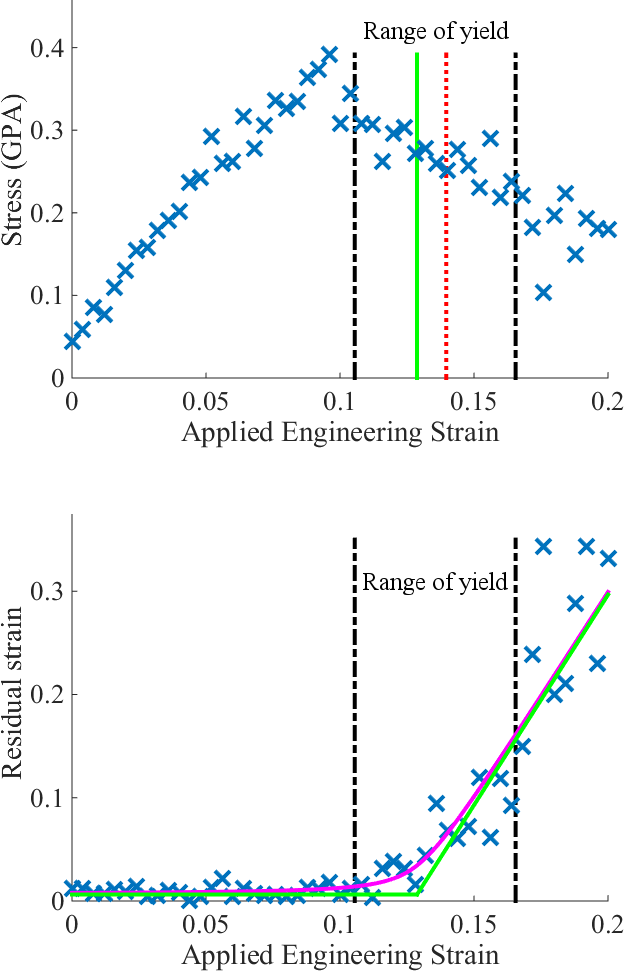}\caption{Example of a dataset 44BA where estimates of yield based on $\sigma(\epsilon)$ differ significantly from the residual strain analysis.  Colors and linestyles for the vertical lines have the same interpretations as in Fig.~\ref{fig:newss}. Interestingly, the peak of the stress-strain data appears well resolved but lies outside of the confidence interval predicted by the hyperbola analysis.}\label{fig:badpeak}
\end{figure}

\subsection{Comparison with experiments}
While our main goal is to propose and {\it verify} a {new procedure for estimating $\epsilon_y$ from MD simulations, our approach would nonetheless be problematic} if it showed {little} correspondence with real experiments.  To this end, we briefly discuss past results that support the usefulness of our method.

\textcolor{black}{In 1996, Quinson {\it et al.} showed that deformation-relaxation experiments can be used to quantify the rate-dependence of relaxation modes in linear polymers such as polystyrene \cite{ref1}.  As a byproduct of this work, they generated plots of residual strain data as a function of the applied strain.  Interestingly, they observed that yield (or the onset of plastic deformation) occurred at the first value of  for which the material exhibited a non-zero residual strain; cf.\ also  Refs.~\cite{Perez98,Pegoretti06,PEN97}.  Although not discussed by the authors, it is noteworthy that in all of their results, $\epsilon_r$ is approximately a linear function of the applied strain beyond yield.  Critically, this observation holds irrespective of either the deformation rate, temperature, or relaxation time.  Given the inherent length and time-scale limitations of MD, these observations are therefore encouraging, since they suggest that our simulated estimates of $\epsilon_y$ are independent of the time over which the system is allowed to relax.  }

Reference \cite{Jackson16} recently reproduced some of these results via strain-relaxation experiments on 44BA.  Certain details of their analysis differ from our approach.  Notably, they computed residual strain relative to the direction of major strain only; moreover, their compression tests were not necessarily volume conserving.  Nonetheless, key elements of their results are replicated in our simulations.  In particular, their residual-strain data exhibits the same bilinear character evident in, e.g.\ our Fig.~\ref{fig:residdata} and Refs.~\cite{ref1,Perez98,Pegoretti06,PEN97}.  Moreover, they identified the onset of permanent deformation with the appearance of a non-zero slope in their $\epsilon_r$.  Such observations are promising in that they suggest that the simulations capture the same phenomena as experiments.  

More direct comparison with the results in Ref.~\cite{Jackson16} is complicated by several factors.  In particular, the authors did not measure the strain perpendicular to the direction of compression; thus it is not compute the Poisson ratio of the deformation.  Moreover, our procedure for creating crosslinked structures has free parameters controlling properties such as the final crosslink percentage, etc.  Thus, while it is likely that our method can be calibrated to {be comparable with experimental measurements}, such an analysis cannot be performed with the currently available information.

\subsection{Limitations and open problems}

As suggested in Sec.~\ref{sec:intro} and Ref.~\cite{Patrone16}, a key feature of small-scale molecular dynamics simulations is the inability for any one realization to sample all of the possible structures that a crosslinked network can form.  In Ref.~\cite{Patrone16}, the authors showed how this undersampling can lead to significant uncertainties that are not captured in the analysis of a single simulation.  They addressed this problem by proposing a weighted-mean statistic to estimate the unaccounted for "dark" uncertainties by comparing results from multiple independent simulations.  In general, we expect that a similar undersampling of the crosslinked network gives rise to varying estimates of yield between simulations.  But while it is desirable to extend the dark-uncertainty analysis to the case at hand, certain problems arise.    

In particular, such an analysis requires that the individual datasets be converged as a function of system size so that the hyperbola analysis does not bias predictions via nonlinear transformations of the noise.  Reference \cite{Patrone16} addressed this problem through a pooling analysis that detected bias by systematically averaging datasets to mimic the effect of reducing noise.  Importantly, however, this analysis was facilitated by the fact that $T_g$ is extracted from scalar relationships (i.e.\ density-temperature curves); thus it was straightforward to formulate the notion of an average density.  In the case of yield-strain, the corresponding raw data is composed of tensors (which are subsequently massaged into scalars).  Given that the underlying unit cells are non-orthogonal, it is not straightforward to formulate what we mean by average stress and strain tensors.  That is, the appropriate method of averaging raw data may depend on the relative orientations of the unit cells, which requires additional modeling or assumptions beyond what we propose.

{\it Acknowledgements} The authors thank Anthony Kearsley, Stephen Christensen, and Matthew Jackson for helpful discussions during preparation of this manuscript.  The authors also thank The Boeing Company for computational resources provided in support of this project.  P.P. was partially funded by the Institute for Mathematics and its Applications (IMA) as a postdoctoral fellow during this work.  The IMA is a National Science Foundation Math Institute funded under award DMS-0931945.

\appendix

\section{Convergence criterion}
\label{app:convergence}

The general method for assessing convergence of our simulated quantities works as follows.  First, let $\rho$ denote an arbitrary quantity such as density, stress, or length.  Denote an interval of times between time $t_j$ and $t_{j+1}$ as $I_j := [t_j,t_{j+1}]$, and partition this interval into $N_j$ equal timesteps $t_{j,k}$.  We use $I_j$ to represent a single simulation whose duration is equal to the length of the interval; $t_{j,k}$ ($1\le k \le N_j$) denote the discrete simulation timesteps within that interval. 

For both the equilibration and $\rho$-averaging algorithms, we first run a single simulation  and compute the running average
\begin{align}
\bar \rho_r(t_{1,k}) = \frac{1}{k} \sum_{k'=1}^k \rho(t_{1,k'}),
\end{align}
where $ \rho(t_{1,k'})$ is the quantity output by the simulation at the $k'$th step in the 1st interval.  Next, we compute the ``variance''
\begin{align}
V_{j} &= \frac{1}{N_{j}} \sum_{k=1}^{N_{j}}[ \bar \rho_r(t_{j,k}) - \bar \rho_r(t_{j,N_{j}})]^2
\end{align}
for $j=1$.  If $V_1$ is less than some user-defined threshold $V_{\rm target}$, then the algorithm stops.  If, on the other hand, $V_1 > V_{\rm target}$, we run additional simulations, computing
\begin{align}
\bar \rho_r(t_{j+1,k}) &= \frac{   \left[\sum_{j,k'}   \rho(t_{j,k'}) \right]  + \sum_{k'=1}^k \rho(t_{j+1,k'})     }{\left[\sum_{j} N_j \right] + k} 
\end{align}
until $V_j < V_{\rm target}$.  Note that we calculate the running average across the sum of intervals $\sum_j I_j$, whereas we only compute the variance $V_j$ for individual intervals $I_j$.

In general, the task of picking suitable interval lengths $I$ and convergence criteria $V_{\rm target}$ is non-trivial.  Degenerate cases illustrate potential problems.  If $I$ is so small as to encompass only a few timesteps, the system will not evolve from its initial state, and the variances $V_j$ will be artificially small.  Consequently, the algorithm will stop prematurely.  In the other extreme, picking $I$ to be on the scale of nanoseconds (which is long for MD simulations) can lead one to do more dynamical averaging than may be necessary.  Picking $V_{\rm target}$ too large or small leads to similar problems.  For condensed polymer systems with $\mathcal O(10^4)$ atoms or fewer, we find that setting $I=20$ ps is a reasonable compromise for all temperatures between 100 K and 800 K, provided that our timesteps are 1 fs.  This yields $N_j = 20,000$ for all $j$.  For the density equilibration and averaging steps, we set $V_{\rm target} = 10^{-3}$ g${}^2$/cm${}^6$ and $V_{\rm target} = 10^{-7}$ g${}^2$/cm${}^6$, respectively.  For stress-equilibration and averaging steps, we set $V_{\rm target}=10^{-2}$ GPa${}^2$ and $V_{\rm target}=10^{-3}$ GPa${}^2$.  For the cell-dimension equilibration and averaging steps, we set $V_{\rm target}=10^{-5}$ nm${}^2$ and $V_{\rm target}=5 \times 10^{-9}$ nm${}^2$.  In each of these cases, convergence means that fluctuations in the corresponding running averages are less than $\pm 3 \sqrt{V_{\rm target}}$.

We caution that this stopping algorithm can severely undersample metastable energy minima of a single system, especially when the characteristic time to leave the mimima is on the order of or greater than the simulation lengths $I_j$.  We attempt to compensate for this problem by generating multiple realizations of the same chemistry in order to more fully explore its possible configurations.  

\textcolor{black}{It is also important to note that our convergence criterion does not guarantee that the system statistics are representative of realistic equilibrium distributions.  Reference \cite{shirts} provides an analysis for assessing the extent to which this latter criterion has been satisfied on MD timescales, and this approach can likely be incorporated into our simulation procedure.  However, we do not explicitly quantify uncertainties due to lack of such convergence because MD cannot reach the timescales necessary to verify that the aforementioned analysis is applicable to crosslinked polymers in the glassy regime. }

\section{Strain control for non-orthogonal unit cells}
\label{app:non-orthogonal}

Our MD simulations model non-orthogonal, periodic unit cells with shapes that vary between realizations.  In order to compare simulation datasets with different underlying structures, it is therefore useful to strain all of the systems in the same way, e.g.\ relative to an orthogonal coordinate system.  In practice, this entails changing both the length of and angles between the non-orthogonal basis vectors.  

To achieve this, we first pick an orthogonal coordinate frame in which to strain the system.  Without loss of generality, we assume that the longest basis vector of the non-orthogonal system coincides with the $z$-axis of its orthogonal counterpart.  In the orthogonal basis, we impose a volume conserving strain 
\begin{align}\epsilon=
\begin{bmatrix}
\lambda & 0 & 0 \\ 0 & \lambda & 0 \\ 0 & 0 & -\Delta
\end{bmatrix} \label{eq:straintensor}
\end{align}
where $\lambda > 0$ and $\Delta > 0$ are small.  Letting $\one$ denote the identity matrix, volume conservation implies that $|\one + \epsilon|=1$, where $|\star|$ denotes the determinant of $\star$.  Consequently, 
\begin{align}
\lambda = \sqrt{\frac{1}{1-\Delta}} - 1,
\end{align}
so that picking the strain increment associated with compression determines expansion in the other two  orthogonal directions.  

To compute the associated changes in the non-orthogonal unit cell, first let $A,$ $B$, and $C$ denote its three unit vectors, recall that $C$ is parallel to the $z$-axis, and assume that $B$ is in the $y$-$z$ plane.  The angles between these unit vectors are defined via the inner products
\begin{align}
C\cdot B &= \cos(\alpha) \\
C \cdot A &= \cos(\beta) \\
B\cdot A &= \cos(\gamma). \label{eq:bdota}
\end{align}
It is convenient to encapsulate this information by writing $A$, $B$, and $C$ in terms of the orthogonal coordinate system; viz.
\begin{align}
C&=\z \\
B&=\sin(\alpha)\y + \cos(\alpha)\z \\
A&=\chi \x + \mathfrak y \y + \cos(\beta)\z
\end{align}
where $\x$, $\y$, and $\z$ are the orthogonal unit vectors, and 
\begin{align}
\mathfrak y &= \frac{\cos(\gamma) - \cos(\alpha)\cos(\beta)}{\sin(\alpha)} \\
\chi &= \sqrt{1-\mathfrak y^2 - \cos(\beta)^2}.
\end{align}
These last identities are computed via the aid of Eq.~\eqref{eq:bdota} and the fact that $|A|=1$ by definition.

Now, straining the system according to Eq.~\eqref{eq:straintensor} amounts to computing new basis vectors $\x'$, $\y'$, and $\z'$ via the relation
\begin{align}
\begin{bmatrix} \x' \\ \y' \\ \z'    \end{bmatrix} = \begin{bmatrix}
1+ \lambda & 0 & 0 \\ 0 &1+ \lambda & 0 \\ 0 & 0 &1 -\Delta
\end{bmatrix}\begin{bmatrix} \x \\ \y \\ \z    \end{bmatrix}.
\end{align}
The non-orthogonal unit vectors become
\begin{align}
C'&=(1-\Delta)\z \\
B'&=(1+\lambda)\sin(\alpha)\y + (1-\Delta)\cos(\alpha)\z \\
A'&=(1+\lambda)\chi \x + (1+\lambda)\mathfrak y \y + (1-\Delta)\cos(\beta)\z.
\end{align}
The lengths of the strained, non-orthogonal basis vectors can be easily calculated from the above.  Moreover, we can define strained angles by inverting the definitions
\begin{align}
C'\cdot B' &= \cos(\alpha') \\
C' \cdot A' &= \cos(\beta') \\
B'\cdot A' &= \cos(\gamma'). \
\end{align}

\bibliography{yield.bib}

\end{document}